# Some further developments on a neurobiologically-based model for color sensations in humans

*Charles Q. Wu; Perception and Cognition Research, Redmond, WA, U.S.A.; Email: charlesqwu@percog.org*

## Abstract

At HVEI-2012, I presented a neurobiologically-based model for trichromatic color sensations in humans, mapping the neural substrate for color sensations to V1-L4—the thalamic recipient layer of the primary visual cortex. In this paper, I propose that V1-L4 itself consists of three distinct sub-layers that directly correspond to the three primary color sensations: blue, red, and green. Furthermore, I apply this model to three aspects of color vision: the three-dimensional (3D) color solid, dichromatism, and ocular agnosticism. Regarding these aspects further: (1) 3D color solid: V1-L4 is known to exhibit a gradient of cell densities from its outermost layer (i.e., its pia side) to its innermost layer (i.e., its white matter side). Taken together with the proposition that the population size of a cell assembly directly corresponds with the magnitude of a color sensation, it can be inferred that the neurobiologically-based color solid is a tilted cuboid. (2) Chromatic color blindness: Using deuteranopia as an example, at the retinal level, M-cones are lost and replaced by L-cones. However, at the cortical level, deuteranopia manifests as a fusion of the two bottom layers of V1-L4. (3) Ocular agnosticism: Although color sensation is monocular, we normally are not aware of which eye we are seeing with. This visual phenomenon can be explained by the nature of ocular integration within V1-L4. A neurobiologically-based model for human color sensations could significantly contribute to future engineering efforts aimed at enhancing human color experiences.

## Terminology

### Two Meanings of the Word "Color"

In everyday language, the word *color* has two distinct meanings: It may or may not include the black–white dimension. James Clerk Maxwell (1831–1879), a pioneer in the experimental study of human color vision, begins a review paper on color vision with the following statement: "All vision is colour vision, for it is only by observing differences of colour that we distinguish the forms of objects. I include differences of brightness or shade among differences of colour." (Maxwell, 1871, p. 13). Christine Ladd-Franklin (1847–1930), one of the first American woman psychologists, expresses the same perspective: "The sadly ambiguous word *colour* should be used in its inclusive sense, as including the whites, black, and the black-whites (or greys). The colour-sensations should then be divided into the achromatic and the chromatic ones . . . " (Ladd-Franklin, 1929, p. 283)

As evident from these statements, both Maxwell and Ladd-Franklin recognize the necessity of including the black-white dimension in the definition of *color*. In this paper, I will follow their perspective—hence, the term *color*, as used here, is not just about *hues*. In this perspective, the statement "white is color-less" is meaningless: White is a special color—more specifically, white is a special combination of all colors.

### Color Sensation, Visual Sensation, and Visual Consciousness

Throughout this paper, I will use the term *sensation* in both singular and plural forms. I will use the latter when referring to various colors. Additionally, I will use the terms *color sensation*, *visual sensation*, and *color consciousness* interchangeably. The rationale for using these terms interchangeably aligns with Maxwell's statement mentioned previously: "All vision is colour vision." Here, *vision* means "seeing at a conscious level"; therefore, Maxwell's statement means that *color sensation* is the basis (or say, carrier) of *visual consciousness*.

Maxwell inherited this notion regarding color vision from his philosophical mentor, Sir William Hamilton (1788–1856). As illustrated in Figure 1, Hamilton conceptualizes the activity of a living brain as consisting of both subconscious and conscious components, with the latter consisting of *sensation* and *perception*. He further claims that, for visual consciousness, *visual sensation* is essentially *color sensation*.

This paper is not a philosophical study. I will neither elaborate nor argue for Hamilton's scheme of conceiving consciousness. However, this scheme can be traced to philosophers who preceded Hamilton. In this regard, interested readers should refer to Hamilton (1858).

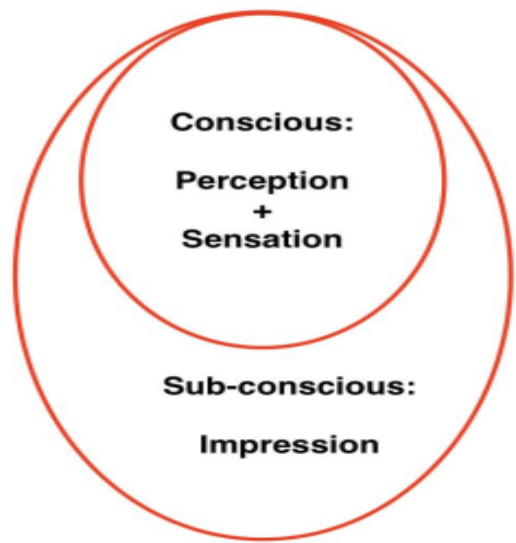

*Figure 1.* Sir William Hamilton's view on consciousness, perception, sensation, and impression.

## Introduction

Broadly speaking, this paper addresses the *neural correlate of consciousness* (NCC) problem (see Crick & Koch, 1995, 2003). In the context of color vision. Judd and Wyszecki (1975) stated the same problem as follows: "Neural correlate of visual surroundings: Does there exist anywhere in the brain a place where a display of the objects in our visual space is built up from nerve signals initiated by the opening of our eyes, and where this display is continually kept current from second to second in response to the myriad signals produced by motion of the objects in our visual space and by our eye, head, and body movements?" (p. 26). The "display" alluded to here has also been referred to as the "Cartesian Theater" (Dennet, 1991). As illustrated in Figure 1, color sensation is part of visual consciousness. Therefore, our question is: Does there exist anywhere in the brain a place for color sensation? The answer that I will offer in this paper is that Layer 4 (i.e., the thalamic recipient layer) of the primary visual cortex (i.e., visual cortical area V1) is that place. Furthermore, I propose that V1 Layer 4 consists of three sub-layers corresponding to the three primary colors, as encapsulated in the Young-Maxwell-Helmholtz trichromatic theory.

This paper is organized as follows: First, I will show that the Opponent-Colors Theory is wrong, a necessary step for the argument that follows through the rest of this paper. Second, I will demonstrate that color sensation is essentially monocular and will then map this characteristic of color sensation onto the neuroanatomical organization of the human visual system. Finally, I will establish a neurobiologically-based model for color sensations and will then use this model to explain some relevant color vision phenomena.

## The Opponent-Colors Theory (OCT) is Wrong!

OCT was first proposed by Ewald Hering in the late nineteenth century (Hering, 1964) and later updated by American psychologists Leo Hurvich and Dorothy Jameson in the mid-twentieth century (Hurvich & Jameson, 1957). Presently, OCT appears in every textbook that touches upon the subject of color vision (e.g., Wolfe et al., 2021, p. 143—This textbook is very well-written and beautifully-illustrated; here it is just used as an example to illustrate the ubiquitousness of OCT in relevant textbooks). However, OCT is fundamentally wrong. Many lines of theoretical, psychological, and neurophysiological evidence have exposed its flaws, either as inconsistencies with experimental data or as logical defects within the theory itself.

First, a core assumption of OCT is the existence of four "unique" hues: RED, GREEN, YELLOW, and BLUE. But is YELLOW a unique hue? Maxwell (1871) experimentally demonstrated that YELLOW is a mixture of GREEN and RED: "Proceeding from this green towards the red end of the spectrum, we find the different colours lying almost exactly in a straight line. This indicates that any colour is chromatically equivalent to a mixture of any two colours on opposite sides of it and in the same straight line" (p. 15).

Maxwell then reflects on the discrepancy between this experimental conclusion and the ordinary experience of perceiving YELLOW as something special or unique, noting "between the red and the green we have a series of colours apparently very different from either, and having such marked characteristics that two of them, orange and yellow, have received separate names. . . . I do not profess to reconcile this discrepancy between ordinary and scientific experience. It only shows that it is impossible, by a mere act of introspection, to make a true analysis of our sensations. Consciousness is our only authority; but consciousness must be methodically examined in order to obtain any trustworthy results" (p. 15). Furthermore, OCT claims that it is impossible to see RED and GREEN in the same "place" and at the same time. However, there has been evidence that under special viewing conditions, it is possible to see RED and GREEN in the same "place" and at the same time (Crane & Piantanida, 1983; Billock & Tsou, 2010).

Second, OCT adherents have incorrectly used opponent color relations (i.e., RED-GREEN, YELLOW-BLUE) to explain complementary color phenomena. Afterimage has always been an intriguing visual phenomenon to me; due to my obsession with it, having observed thousands of complementary afterimages under various illuminating conditions, I realized (or say, re-discovered) in 2008 that the complementary color to RED is not GREEN, but CYAN, and that to GREEN is not RED but MAGENTA. When I began looking into the color vision literature of the nineteenth century, I was quite surprised to see that these complementary color relations were already well-known during Hering's lifetime. Ladd-Franklin (1899, 1929) had clearly articulated these relations and stridently opposed Hering's evocation of opponent colors to explain complementary color phenomena.

Searching deeper into the color vision literature before Hurvich and Jameson, I also encountered William McDougall's work on vision (McDougall, 1901a, 1901b, 1901c, 1911), which, unfortunately, still remains overlooked as of now. My observations of complementary afterimages, combined with what I learned from Ladd-Franklin and McDougall, had led me to reject OCT and denounce it as a completely untenable theory (Wu, 2009a, 2009b, 2010, 2012a).

Around the same time, Pridmore (2008, 2011) also argued that complementary colors, rather than opponent colors, underlie many color phenomena. However, he still kept an opponent-colors stage in his model for color vision. Additionally, veteran vision scientist Donald MacLeod (2010) published an insightful book chapter, where he did not explicitly denounce OCT, but did place OCT under deep scrutiny. In recent years, more and more vision researchers have questioned and/or criticized OCT: For example, Tyler (2020) confirmed the complementary color relations exhibited in complementary afterimages, and Conway et al. (2023) directly announced "the end of Hering's Opponent-Colors Theory".

When Young (1801) was formulating the trichromatic theory, he was indeed intending this theory for color sensations. But at that time, he was also making an implicit assumption: The visual brain somehow mirrors the retina. This assumption was later formally proposed by Johannes Müller as the Law of Specific Nerve Energies (see Riese & Arrington, 1963). However, now we know that in the visual modality there is a transformation from three cone photoreceptors' spectral selectivities to the three primary color sensations (Pridmore, 2011, Fig. 5). Furthermore, as remarked by MacLeod (2010), "[R]ods provide a fourth visual pigment and hence a fourth degree of freedom for the neural effects of color stimuli. Rods are important for color vision under a wide range of illumination levels. When rods as well as cones are involved, trichromacy is not established at the photoreceptor level" (p. 155). In these respects, the Law of Specific Nerve Energy does not completely hold for the neural pathway from the very beginning of the visual modality (i.e., photoreceptors, which are 4-dimensional as claimed by MacLeod) to color sensations (which are 3-dimensional, as conceived by Young and evidenced by substantial color-matching data accumulated by Maxwell and others). We should still uphold the trichromatic theory as a theory for color sensations (see Maxwell's emphasis on sensation and consciousness in Maxwell, 1871). If we believe in this, we will be able to find—and indeed, to easily find—the neural correlate for color sensations. On the other hand, if we continue adhering to OCT, we will not be able to find any brain structure whose anatomical and physiological properties match to what OCT conceives.

## Color Sensation is Monocular

Now we come to an important characteristic of color sensation—that is, its monocularity; in other words, color sensation must occur at a monocular stage within our visual system. This notion is not new: Many researchers have advocated it. Helmholtz (1867/1925) characterizes this notion as follows: "… the content of each separate field [from one eye] comes to consciousness without being fused with that of the other field by organic mechanisms; and that, therefore, the fusion of the two fields in one common image, when it does occur, is a psychic act" (p. 499). Sherrington (1904) and McDougall (1911) later picked up this notion and further developed it.

Here I will use one fascinating visual phenomenon to demonstrate the monocular characteristic of visual sensation. As we know, there is a blind spot (BS) in each of our eyes. The BS was discovered by Edme Mariotte around 1668; Mariotte's method of demonstrating the BS, however, is about how to map it within one's visual field, not about how to (consciously) see it. Presently, any textbook in psychology, neuroscience, and ophthalmology, when teaching about the BS, teaches this method only (e.g., Wolfe et al., 2021, p. 40).

Under special conditions, it is actually possible to see one's own BS in each eye, literally seeing the BS as a black hole on a lighter background or a white hole on a darker background—more generally, a colored spot on a background of the spot's complementary color; the BS may, or may not, be accompanied by the Purkinje Tree (PT) which denotes the image of retinal blood vessels. As far as I have been able to trace back, this phenomenon was first reported by the French scholar Philippe de La Hire (1640-1718) in La Hire (1694). Figure 2 is La Hire's drawing of his entoptic vision when he saw the BS, the PT, and the central blue scotoma (caused by the absence of blue cones in the foveal region of the retina; under special viewing conditions, this region may show up in one's visual field–-see, Magnussen et al., 2004). I have previously suggested that this phenomenon be named the La Hire phenomenon (Wu, 2021b). Over history, the La Hire phenomenon has been re-discovered again and again—see Helson (1929, pp. 352–353), Brøns (1939, Chapter IV), and Wu (2021b) for descriptions of such re-discoveries. In 2012, I serendipitously re-discovered it too and subsequently presented it at that year's annual meeting of the Society for Neuroscience (Wu, 2012b). Strangely, this phenomenon has never entered textbooks and mainstream knowledge—The only textbook that I have found to contain a description of this phenomenon is Titchener (1911), as he relates: "It has recently been asserted that if one looks suddenly, with a single eye, at some uniform and brightly illuminated surface, one sees the projection of the blind spot as a faint grey patch" (p. 328). In recent years, I have been attempting to popularize this wonderful visual phenomenon (Wu, 2021b, 2024).

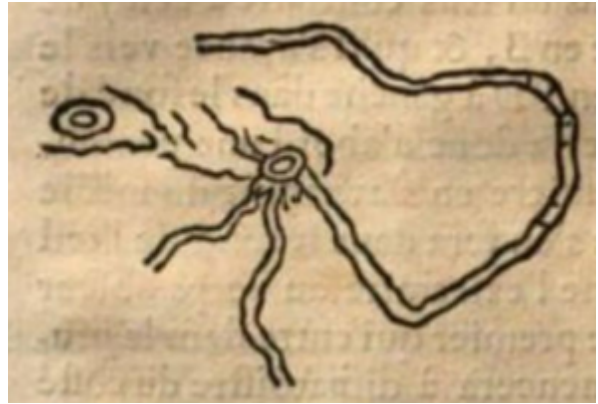

*Figure 2.* La Hire's drawing of his entoptic vision: Seeing the BS (the spot at the center) of his left eye, PT (Purkinje Tree; the image of retinal blood vessels, depicted as branches coming out from the BS), and the Central Blue Scotoma (the spot on the left).

Since the BS in each eye is specific to that eye, the La Hire phenomenon clearly indicates that visual sensation (i.e., color sensation) is monocular.

## Mapping the Blind Spot in the Brain

The early stages of the human (or more broadly, the primate) visual system consists of the retina, the lateral geniculate nucleus (LGN) of the thalamus, and the primary visual cortex (i.e., V1). The cortical sheet comprises six layers, with Layer 4 receiving thalamic inputs (i.e., optic radiations in the case of V1). Hereafter, we will denote this layer as V1-L4. Please note that "Layer 4" in V1 has been incorrectly labeled as "Layer 4C" in many textbooks (e.g., Wolfe et al., 2021, p. 71); see Boyd et al. (2000) and Balaram et al. (2014) for the relevant neuroanatomical evidence as to why it should be labeled as "Layer 4" instead of "Layer 4C".

The La Hire phenomenon is a wonderful psycho-anatomical means: As a matter of fact, several neuroanatomical studies have precisely localized a representation of the BS in V1-L4: Figure 3 shows two such studies (LeVay et al., 1985; Adams et al., 2007), with the first one on the monkey brain and the second on a human brain. Though in two species and using different chemical staining methods, both studies clearly indicate that there is a representation of the BS in V1-L4. Please note that V1-L4 is a "bi-monocular" structure in the sense that for each and every tiny patch of the viewer's binocular visual field, the monocular image (i.e., ocular dominance column or ODC) from one eye resides with that for the other eye side by side. For the upper slide in Figure 3, white stripes and areas depict neural tissue regions in V1-L4 predominantly connected with the eye containing the BS, whereas the black stripes and areas depict that connected with the other; for the lower slide in Figure 3, the labeling of the white and the black stripes / areas is just the opposite. From these slides, we should understand that the representation of the BS in V1-L4 does not create any physical "hole" in this neural tissue—instead, the area is invaded and occupied by the input from the other eye.

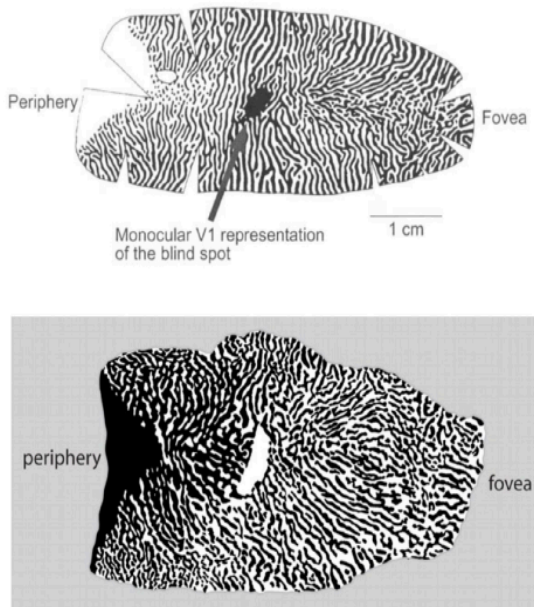

*Figure 3.* Two neuroanatomical studies show representations of blind spots in V1-L4 (i.e., the thalamic recipient layer in the primary visual cortex).

Beyond V1-L4, is there any other neural structure(s) in the primate visual system that may contain representations of the BS? David Hubel and Torsten Wiesel's pioneering exploration of the feline and the primate visual brains had long established that neurons in V1-L4 are primarily monocular whereas that beyond V1-L4 are mainly binocular (e.g., Hubel & Wiesel, 1968). As we already stated, each eye's BS is specific to that eye (i.e., monocular); therefore, the answer to this question is negative.

Correlating the La Hire phenomenon with such neuroanatomical studies, we can conclude that visual sensation is represented in V1-L4. Please note that without knowing the La Hire phenomenon, we cannot argue that the BS representations seen in V1-L4, and this layer in general, are directly correlated with visual sensations—in other words, one may argue that such representations are just for sub-consciousness neural activation; then, with the knowledge of the La Hire phenomenon, we can indeed pinpoint the neural substrate for visual sensations to V1-L4. (The authors of the relevant neuroanatomical studies did not link their results to any visual phenomenon. Therefore, my contribution here is conceptual in linking neuroanatomical studies, on the one hand, with the La Hire phenomenon on the other.)

## Neurobiologically-based Model for Color Sensations

Textbooks (e.g., Wolfe et al., 2021, p. 70) introducing the neuroanatomical organization of the primate visual system usually teaches the following geniculo-cortical wiring scheme: For each retina, the M-layer (magnocellular) in the LGN projects to layer 4Cα in V1; and the two P-layers (parvocellular) project to layer-4Cβ. But, why does our Lord (or say, Mother Nature) twist two bundles of neural fibers together into one on the geniculo-cortical route? Here, the following saying from Sir Isaac Newton (1643–1727) may be useful as a general guidance when considering biological structures: "Nature does nothing in vain. … for God in the frame of animals has done nothing without reason" (Newton, 1682/1850, p.270).

In the words of Boyd et al. (2000), "...not recognizing it [i.e., a third sub-layer in V1-L4] may have led to errors of interpretation in previous studies which placed data on connections of three anatomical divisions into two conceptual compartments" (p. 645). Indeed, recently there has been neuroanatomical evidence (Boyd et al., 2000, Fig. 1, p. 655) indicating that there are three sub-layers within V1-L4—this naturally fits with Young-Maxwell-Helmholtz's trichromatic theory. Here I propose that the organizational feature of three divisions per retina in the LGN is still conserved within V1-L4—though there may be a transform from the three cone-based (i.e., S-, M-, and L-cones) spectral selectivity functions in the LGN to the three primary color sensations (i.e., BLUE, RED, and GREEN) in V1-L4. Another prominent neuroanatomical feature of V1-L4 is that there is a gradual increase of cell density from the top of the layer to its bottom (see Lund et al., 1995).

Mapping onto Young-Maxwell-Helmholtz's trichromaticity for color sensations, we have the following six-pack model (more formally, a sexpartite model) for V1-L4: Tangentially, it consists of ODCs receiving thalamic inputs from the two retinas respectively; Vertically, from the top of the layer (i.e., the pia side) to its bottom (i.e., the white-matter side), the layer itself consists of three sub-layers corresponding to the three primary color sensations: BLUE, GREEN, and RED. This six-pack model for V1-L4 is illustrated in Figure 4; since our emphasis in the present paper is on the cortical stage for color sensation, here we will leave open the exact transformation from the retinal 4-dimensional spectral response space to the 3-dimensional perceptual color space in V1-L4.

Perceptually, under the same illumination, BLUE appears the least bright, and GREEN the most; RED is close to GREEN but still less than GREEN in brightness. There have been quantitative data substantiating this perceptual experience—e.g., Pitt (1944, Figure 7), Graham and Hsia (1958, Figure 7). This is why the three primary colors are arranged in the order of BLUE, RED, and GREEN in Figure 4. The neurons in V1-L4 have overlaps in their dendrites across the three-layers (see Lund et al., 1995), and such a neural wiring can explain a peculiarity in the spectral selectivities of the three primary colors—that is, RED appears at both the short-wave and the long-wave ends of the visible spectrum. Kuehui (2002) mentioned this peculiarity as "another interesting question" yet to be answered, Oh and Sakata (2015) referred to it as an "incompletely solved mystery", and Pridmore (2016) presented an answer in terms of a plausible neural wiring diagram. The explanation offered here is intimately based on the relevant neuroanatomical features of V1-L4.

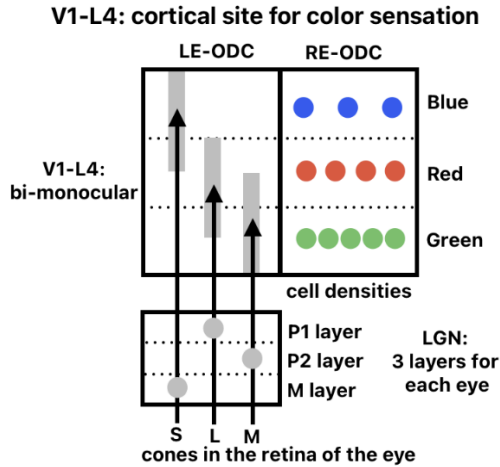

*Figure 4. Neurobiologically-based model for color sensations in the human visual system. LE: left eye; RE: right eye; ODC: ocular dominance column.*

### *The Uniqueness of a Mixture of Two Colors*

The model in Figure 4 offers a new view for understanding a color's "uniqueness"—that is, the uniqueness of a mixture of two colors is related to the two colors' cell densities. YELLOW (as a mixture of RED and GREEN), looks unique because both RED and GREEN sensations have high cell densities in V1-L4; on the other hand, PURPLE (as a mixture of RED and BLUE), exhibits granularities of its mixture components because BLUE sensation has much low cell density in V1-L4. This new view about color uniqueness constitutes another blow to Hering's OCT—in addition to those already advanced by Ladd-Franklin (1899), McDougall (1901a, 1901b, 1901c, 1911), Wu (2012), and Conway et al. (2023).

### *The Color Space in the Brain is a Tilted Cuboid*

As shown in Figure 4, there are three sub-layers in V1-L4; and these sub-layers possess different cell densities. These two neuroanatomical aspects can be used to establish that a neurobiological-based color space (also known as, "color solid") must be a tilted cuboid, as illustrated in Figure 5.

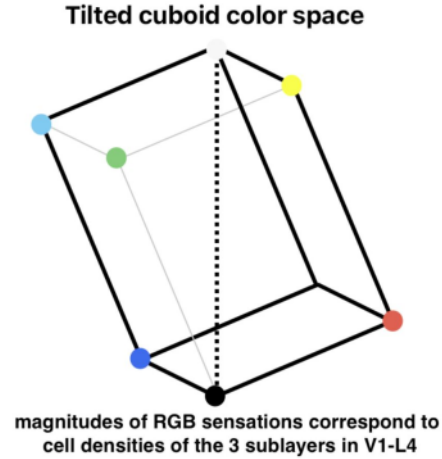

*Figure 5. Neurobiologically-based color space as a tilted cuboid.*

Historically, many schemes for ordering colors have been proposed (see Kuehni, 2003). The color-ordering scheme in the form of a tilted double pyramid offered first by Hermann Ebbinghaus in 1902 and later by Edward B. Titchener in 1909 is very close to the tilted cuboid color solid here (Kuehni, 2003, pp. 81–82).

As remarked by Mollon (2003), Newton's construction of his color circle is a stroke of genius: Two prominent properties of colors manifest in the color circle are circularity (i.e., the notion that all colors can be arranged on a circle) and complementarity (i.e., the notion that two colors on the opposite sides of the color circle are complementary). I first learned about Newton's color circle from the late Prof. Herbert A. Simon (1916–2001), and he and I discussed the problem as to whether circularity causes complementarity or vice versa. The color solid in Figure 5 offers a clear answer to this problem: Color circularity is due to the fact that all colors need to be represented within a neuroanatomical enclosure (i.e., V1-L4), and complementarity naturally follows circularity.

### *Neural Code for Color Sensation*

Implicit in the compound-color theory, there is the following neurophysiological problem: For a single perceivable point (or say, "pixel") in the visual field, how does the brain create a whole gamut of perceivable colors from a set of primary colors? As mentioned above, Helmholtz had referred to this neural process by "psychic fusion" (Helmholtz, 1925, p. 499); but he had never substantiated this psychic process in terms of any conceivable underlying neural mechanism.

In computational neuroscience, the problem of combining features along multiple dimensions to create a coherent representation over a distributed neural network is known as the "binding problem". It has long been suggested that the solution to the binding problem is neural synchronization (von der Malsburg, 1994), and there has been a growing body of neurophysiological evidence supporting this conclusion (Engel & Singer, 1991). In color vision, here we have the same type of binding problem—that is, the color fusion problem stated above. In this regard, I suggest that single-moment neural synchronization is indeed the neurophysiological mechanism (i.e., the neural code) for combining outputs from primary color cells to create a whole gamut of color sensations. In this conception, the numbers of cells firing together in a single moment, instead of their firing rates averaged over certain time windows, are the neurophysiological measures of color components and constitute the weights in the color mixing equation: $C = \alpha R + \beta G + \gamma B$, $\alpha + \beta + \gamma = 1$, where C is any color; R, G, and B are the three primary colors; $\alpha$, $\beta$, and $\gamma$ are the weights for color C along the color cardinals. Under this conception, the color matching functions as initially developed and measured by Maxwell (1960), and later fine-tuned and collected at scale by many others (see Judd & Wyszecki, 1975), are the properties of V1-L4, not the properties of the retina.

Furthermore, in this view, for any snapshot of visual consciousness, the bindings at various levels—among spatial points, within one primary color channel, across color channels (i.e., color fusion or mixture), and among visual features—are all due to the same neurophysiological mechanism (namely, single-moment synchronization). This view regarding color sensation as the basis (or say, "carrier") of visual consciousness is illustrated in Figure 6.

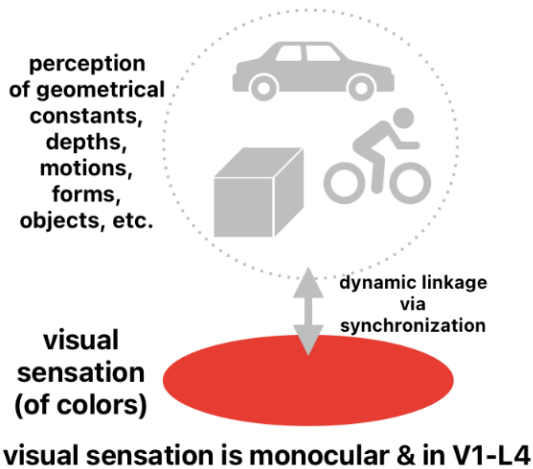

*Figure 6.* Visual sensation (i.e., color sensation) is the basis of visual consciousness. It must happen early in the human visual system.

### Chromatic Color Blindness

This model can also easily explain some peculiarities of chromatic color blindness. For instance, we know that lacking M-cones in the retina would produce deuteranopia and that lacking L-cones would produce protanopia; in terms of color experience, people with these two types of chromatic color blindness both have color sensations along the yellow-blue dimension. As shown in Figure 7, this can be explained by a fusion, or non-differentiation, between the red and the green channels within V1-L4.

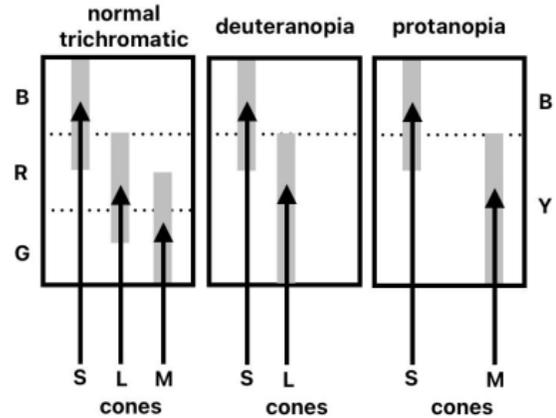

*Figure 7.* Possible V1-L4 configurations explaining trichromatic vision and color deficiencies (e.g., deuteranopia and protanopia).

### Ocular Agnosticism (Blindness to Eye-of-Origin)

A viewer with normal binocular vision is usually not aware of the eye-of-origin information in their visual consciousness. On this point, Helmholtz (1925) writes: "...without making a special experiment for that very purpose [i.e., to know the eye-of-origin information in a normal binocular viewing scenario], we are ignorant as to which image belongs to one eye, and which to the other eye" (p. 459). Since Helmholtz's time, there have been numerous psychological experiments confirming Helmholtz's assertion (e.g., Pickersgill, 1961). We may refer to this visual phenomenon as "ocular agnosticism" (or informally, "blindness to eye-of-origin").

Ocular agnosticism can be explained by the possibility that neural activity in V1-L4 and its associated "location" (i.e., whether in one or the other eye's ODC) are two separate pieces of information; what enters visual sensation may just be the first piece. Furthermore, as illustrated in Figure 8, there may be an ocular "metamerism" in V1-L4: What matters for visual sensation is the total activation from the two eyes' ODCs; how this total activation is distributed in them does not matter.

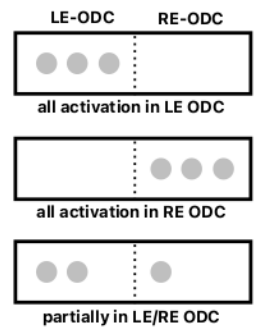

*Figure 8.* Ocular "metamerism": For a particular point in the visual field, what matters for visual sensation is the total activation from LE & RE ODCs. LE: left eye; RE: right eye; ODC: ocular dominance column.

Crick and Koch (1995) mentioned this phenomenon—in their words, "most people are certainly not vividly and directly aware of which eye they are seeing with (unless they close or obstruct one eye) …" (p. 123). This is essentially the same as Helmholtz's statement mentioned above; however, they then incorrectly utilized this phenomenon to argue for their postulate that visual consciousness should happen at a binocular stage. This inference is invalid because that the information for the content of visual consciousness and the eye-of-origin information can be completely independent from each other. (On this matter, I communicated my point to Dr. Crick and he did accept it; see Wu & Crick, 1997).

Another misconception concerning eye-of-origin is a normal binocular viewer's blindness to the change-of-eye-of-origin (hereafter, COEOO). An example of our blindness to COEOO is as follows: In a binocular rivalry setting, when a naïve observer receives different visual stimuli in their two eyes, they can be aware of the recurring changes between the two percepts but be totally unaware of the fact that such changes are due to dichoptic stimulation. I introduced the concept of blindness to COEOO in Wu (2021a); coincidentally, veteran vision scientist Randolph Blake vividly recounted his experience of such blindness as he was sitting through a binocular rivalry experiment for the first time in 1967: "With me seated in front of a peculiar optical device with a separate viewing port for each eye, Fox [i.e., Prof. Robert Fox at Vanderbilt University] illuminated the pictures in the two ports and

Surprisingly the percept stayed stable. This means that the conscious percept stayed stable and at the same time the primary input to layer 4, which is the input layer, in the visual cortex changed. Therefore layer 4 cannot be a part of the neural correlate of consciousness". This inference is incorrect because it fails to take into consideration the observer's blindness to COEOO (Wu, 2021a).

Logothetis et al. (1996) is an ingenious study by incorporating an eye-swapping procedure into the traditional binocular rivalry experimental paradigm. Their experimental result seems to indicate that visual consciousness happens at some binocular stage(s). In the framework illustrated in Figure 6, we can interpret their result as follows: Switching of color sensations occurs in V1-L4 every 333 ms whereas switching of pattern percepts occurs at a binocular stage every 2 ~ 3 s (as illustrated in Figure 9; see also Wu, 1997, 2005). Of course, this interpretation awaits to be tested experimentally.

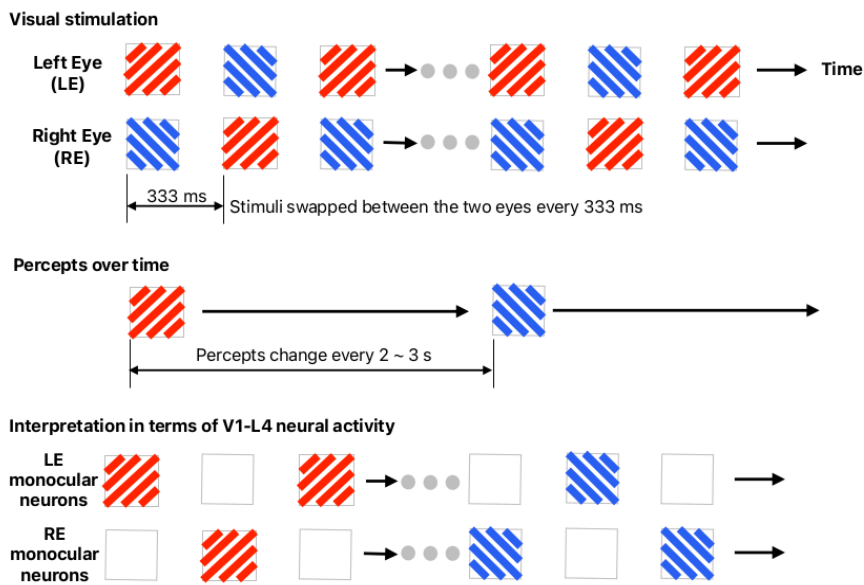

*Figure 9. Interpretation of Logothetis et al. (1996) experimental result: Color sensations occur in V1-L4 and change every 333 ms; pattern percepts occur at binocular stage(s) and change every few seconds.*

asked me what I saw. Unaware that my two eyes were being shown different pictures, I opined that he was showing me a sequence in which the picture of a tree dissolved into the picture of a person's face, and vice versa, over time" (Blake, 2022, p. 139).

Currently on Wikipedia, under the subject matter "neural correlates of consciousness", there is the following inference: "Logothetis and colleagues (i.e., Logothetis et al., 1996) switched the images between eyes during the percept of one of the images.

## Science and Engineering in Color Vision Research

The annual Electronic Imaging Symposium is an excellent venue where industry and academia meet to advance imaging science and applications. At EI-2025, I had been asked several times how my neurobiologically-based model for human color sensations could be applied to electronic imaging. First of all, let me remind us all that in color vision research, science and engineering have always been closely intertwined. Maxwell is an illustrative example in this regard: He is a pioneer in experimentally studying human color vision, including studying color blindness; and he is also an early explorer of color photography.

Second, I would like to quote the following paragraph from Judd and Wyszecki (1975) as a general answer to this question: "The subjective aspect of the nerve activities in the cortex is the color perception itself and is the concern of psychology. … Yet this is the aspect of color in which industry finds its primary concern. All the color control in industry comes down to just one thing—what will the customer see?" (p. 28).

Lastly, I have claimed that the neurobiologically-based color space is a tilted cuboid; as a potential application of my model to color production in industry, will this notion be useful for engineers to expand the gamut of display colors as much as possible to match this physiological color space?

## Conclusions

In Wu (2012a), I made the following claims: (1) The Opponent-Colors Theory is wrong; and (2) V1-L4 is the brain structure where color sensations are produced and represented. In this paper, I have further developed the second claim into the following: V1-L4 consists of three sub-layers, arranged from its top (i.e., the pia side of the cortical sheet) to its bottom (i.e., the white matter side), corresponding to the three primary color sensations in this order: BLUE, RED, and GREEN. These claims form the central tenets of a neurobiologically-based model for human color sensations. To demonstrate this model's explanatory power, I have applied it to several aspects of color vision, including 3D color solid, dichromatism, and ocular agnosticism.

However, I have only barely touched upon how this model may be applied to imaging technologies, an area worthy of exploration. I firmly believe that this model offers a new perspective for future scientific research on color vision as well as further development of imaging technologies to enhance human color experiences.

## Acknowledgements

On November 1, 2024, I presented part of this paper (concerning consciousness and the blind spot) at a research meeting in the Department of Psychology at Carnegie Mellon University. I am grateful to Profs. John R. Anderson, Lynne Reder, and Christian Lebiere for their insightful comments. This paper was accepted by this year's Human Vision and Electronic Imaging (HVEI-2025) Conference and was subsequently selected as a Highlights Session presentation at the Electronic Imaging (EI-2025) Symposium. I am grateful to Dr. Bernice E. Rogowitz (@Visual Perspectives Research and Consulting), Dr. Robin Jenkin (@Nvidia), and the other HVEI-2025 and EI-2025 chairs for their support of this paper. My sincere gratitude also goes to the many attendees at EI-2025 who discussed with me about human visual perception. Last but not least, I am grateful to the late Prof. Herbert A. Simon (1916–2001) for discussing Newton's color circle with me and to the late Dr. Francis Crick (1916–2004) for archiving our correspondences.

## Author Biography

Charles Quanfeng Wu received his MS in electrical engineering from Nanjing Automation Research Institute (NARI) in China (1988) and his PhD in cognitive psychology from Carnegie Mellon University (1992). From 1992 to 1995, he was a postdoctoral researcher in visual neuroscience at the Institute of Ophthalmology, University College London. He conducts independent research in color vision and binocular vision, primarily from theoretical and computational perspectives.